\documentclass[epj]{svjour}
\usepackage{graphics}
\def\beq{\begin{equation}}
\def\eeq{\end{equation}}
\def\beqar{\begin{eqnarray}}
\def\eeqar{\end{eqnarray}}
\def\barr#1{\begin{array}{#1}}
\def\earr{\end{array}}
\def\bfi{\begin{figure}}
\def\efi{\end{figure}}
\def\btab{\begin{table}}
\def\etab{\end{table}}
\def\bce{\begin{center}}
\def\ece{\end{center}}

\def\text{\textstyle}

\def\al{\alpha}

\def\de{\delta}

\def\si{\sigma}
\def\Si{\Sigma}
\def\Ga{\Gamma}
\def\De{\Delta}

\def\refeq#1{\mbox{eq.~(\ref{#1})}}
\def\refeqs#1{\mbox{eqs.~(\ref{#1})}}
\def\reffi#1{\mbox{Fig.~\ref{#1}}}

\def\refta#1{\mbox{Table~\ref{#1}}}

\def\citere#1{\mbox{Ref.~\cite{#1}}}
\def\citeres#1{\mbox{Refs.~\cite{#1}}}

\newcommand{\GeV}{\unskip\,\mathrm{GeV}}

\def\mathswitchr#1{\relax\ifmmode{\mathrm{#1}}\else$\mathrm{#1}$\fi}

\newcommand{\PW}{\mathswitchr W}
\newcommand{\PZ}{\mathswitchr Z}

\newcommand{\PH}{\mathswitchr H}

\newcommand{\Pb}{\mathswitchr b}

\newcommand{\Pt}{\mathswitchr t}

\def\mathswitch#1{\relax\ifmmode#1\else$#1$\fi}

\newcommand{\MW}{\mathswitch {M_\PW}}

\newcommand{\MZ}{\mathswitch {M_\PZ}}
\newcommand{\MH}{\mathswitch {M_\PH}}

\newcommand{\Mb}{\mathswitch {m_\Pb}}
\newcommand{\Mt}{\mathswitch {m_\Pt}}
\newcommand{\scrs}{{}}
\newcommand{\sw}{\mathswitch {s_{\scrs\PW}}}

\newcommand{\cw}{\mathswitch {c_{\scrs\PW}}}
\newcommand{\sweff}{\sin^2 \theta_{\mathrm{eff}}}

\newcommand{\GF}{\mathswitch {G_\mu}}

\newcommand{\mt}{\Mt}

\newcommand{\mb}{\Mb}

\newcommand{\tsf}{\theta\kern-.20em_{\tilde{f}}}
\newcommand{\tsfp}{\theta\kern-.20em_{\tilde{f}\prime}}
\newcommand{\tsq}{\theta\kern-.15em_{\tilde{q}}}

\newcommand{\mf}{m_f}

\newcommand{\lsim}
{\;\raisebox{-.3em}{$\stackrel{\displaystyle <}{\sim}$}\;}

\newcommand{\alps}{\alpha_{\mathrm s}}

\newcommand{\VL}{\left( \begin{array}{c}}
\newcommand{\VR}{\end{array} \right)}
\newcommand{\ML}{\left( \begin{array}{cc}}
\newcommand{\MLd}{\left( \begin{array}{ccc}}
\newcommand{\MLv}{\left( \begin{array}{cccc}}
\newcommand{\MR}{\end{array} \right)}

\newcommand{\tev}{\,\, \mathrm{TeV}}
\newcommand{\gev}{\,\, \mathrm{GeV}}
\newcommand{\mev}{\,\, \mathrm{MeV}}

\newcommand{\BC}{\begin{center}}
\newcommand{\EC}{\end{center}}
\newcommand{\BE}{\begin{equation}}
\newcommand{\EE}{\end{equation}}
\newcommand{\BEA}{\begin{eqnarray}}
\newcommand{\BEAnn}{\begin{eqnarray*}}
\newcommand{\EEA}{\end{eqnarray}}
\newcommand{\EEAnn}{\end{eqnarray*}}
\newcommand{\non}{\nonumber}
\newcommand{\id}{{\rm 1\kern-.12em
\rule{0.3pt}{1.5ex}\raisebox{0.0ex}{\rule{0.1em}{0.3pt}}}}

\def\draftdate{\relax}
\def\mda{\relax}
\def\mua{\relax}
\def\mla{\relax}
\def\draft{
\def\thtystars{******************************}
\def\sixtystars{\thtystars\thtystars}
\typeout{}
\typeout{\sixtystars**}
\typeout{* Draft mode!
         For final version remove \protect\draft\space in source file
*}
\typeout{\sixtystars**}
\typeout{}
\def\draftdate{\today}
\def\mua{\marginpar[\boldmath\hfil$\uparrow$]%
                   {\boldmath$\uparrow$\hfil}%
                    \typeout{marginpar: $\uparrow$}\ignorespaces}
\def\mda{\marginpar[\boldmath\hfil$\downarrow$]%
                   {\boldmath$\downarrow$\hfil}%
                    \typeout{marginpar: $\downarrow$}\ignorespaces}
\def\mla{\marginpar[\boldmath\hfil$\rightarrow$]%
                   {\boldmath$\leftarrow $\hfil}%
                    \typeout{marginpar:
$\leftrightarrow$}\ignorespaces}
\def\Mua{\marginpar[\boldmath\hfil$\Uparrow$]%
                   {\boldmath$\Uparrow$\hfil}%
                    \typeout{marginpar: $\Uparrow$}\ignorespaces}
\def\Mda{\marginpar[\boldmath\hfil$\Downarrow$]%
                   {\boldmath$\Downarrow$\hfil}%
                    \typeout{marginpar: $\Downarrow$}\ignorespaces}
\def\Mla{\marginpar[\boldmath\hfil$\Rightarrow$]%
                   {\boldmath$\Leftarrow $\hfil}%
                    \typeout{marginpar:
$\Leftrightarrow$}\ignorespaces}
\overfullrule 5pt
\oddsidemargin -15mm
\marginparwidth 29mm
}

\begin{document}
\title{Higher-Order Results in the Electroweak Theory}
\author{Georg~Weiglein
}                     
%
%
\institute{Institute for Particle Physics Phenomenology, University of
Durham, Durham DH1~3LE, UK}
\date{Received: date / Revised version: date}
%
\abstract{
The present status of higher-order results in the electroweak theory is
summarised, with particular emphasis on recent two-loop results for the
prediction of the W-boson mass in the Standard Model and leading three-loop
corrections to the rho parameter. The remaining theoretical
uncertainties in the prediction for the W-boson mass and the effective
weak mixing angle are discussed.
} 
\maketitle
\section{Introduction}
\label{intro}

By comparing the experimental results for the electroweak precision
observables, most prominently the W-boson mass, $\MW$, and the effective
weak mixing angle at the Z-boson resonance, $\sweff$, 
with the predictions of the Standard
Model (SM) and extensions of it, the electroweak theory can be tested at
the quantum level.
The current experimental errors in the determination of 
$\MW$ and $\sweff$ are $\de\MW^{\rm exp} = 34$~MeV and 
$\de\sweff^{\rm exp} = 0.00016$~\cite{datasummer2003},
corresponding to a relative accuracy of
0.04\% and 0.07\%,
respectively.

The prediction for $\MW$ is obtained by using as input the Fermi
constant measured in muon decay, $\GF$, the Z-boson mass, $\MZ$, and the 
fine structure constant according to the relation
\beq
\MW^2 \left(1 - \frac{\MW^2}{\MZ^2}\right) =
\frac{\pi \al}{\sqrt{2} \GF} \left(1 + \De r\right),
\label{eq:delr}
\eeq
where the quantity $\De r$ summarises the radiative corrections.
This is done by an iterative procedure, since $\De r$ itself depends on
$\MW$, $\De r = \De r (\MW, \MZ, \MH, \mt, \dots)$.

The effective weak mixing angle at the Z-boson resonance, $\sweff$, is
defined by the effective vector and axial vector couplings for an
on-shell Z~boson, 
\beq
\sweff = \frac{1}{4} \left(1 -
\frac{{\rm Re} \, g_V}{{\rm Re} \, g_A}\right) .
\label{eq:sweff}
\eeq


\section{Higher-order results for $\MW$ and $\sweff$}

The one-loop result for $\De r$~\cite{sirlin} can be written as
\beq
\De r^{(\al)} = \De \al - \frac{\cw^2}{\sw^2} \De\rho +
\De r_{\mathrm{rem}}(\MH),
\label{eq:delrol}
\eeq
where $\cw^2 = \MW^2/\MZ^2$, $\sw^2 = 1 - \cw^2$. It involves large 
fermionic contributions from the shift in the fine structure
constant due to light fermions, $\De\al \propto \log \mf$,
and from the leading contribution to the $\rho$~parameter, $\De\rho$.
The latter is quadratically dependent on the top-quark mass, $\mt$, as a 
consequence of the large mass splitting in the isospin doublet~\cite{velt}.
The remainder part, $\De r_{\mathrm{rem}}$, contains
in particular the dependence on the Higgs-boson mass, $\MH$.
Higher-order QCD corrections to $\De r$ are known at ${\cal O}(\al
\alps)$~\cite{qcd2} and ${\cal O}(\al \alps^2)$~\cite{qcd3,qcd3light}.

Recently the full electroweak two-loop result for $\De r$ has been
completed. It consists of the fermionic
contribution~\cite{2lferm,2lfermb,2lfermc}, which involves
diagrams with one or two closed fermion loops, and the purely bosonic
two-loop contribution~\cite{2lbos}.

\begin{figure}[t]
\vspace{2em}
\resizebox{0.5\textwidth}{!}{%
  \includegraphics{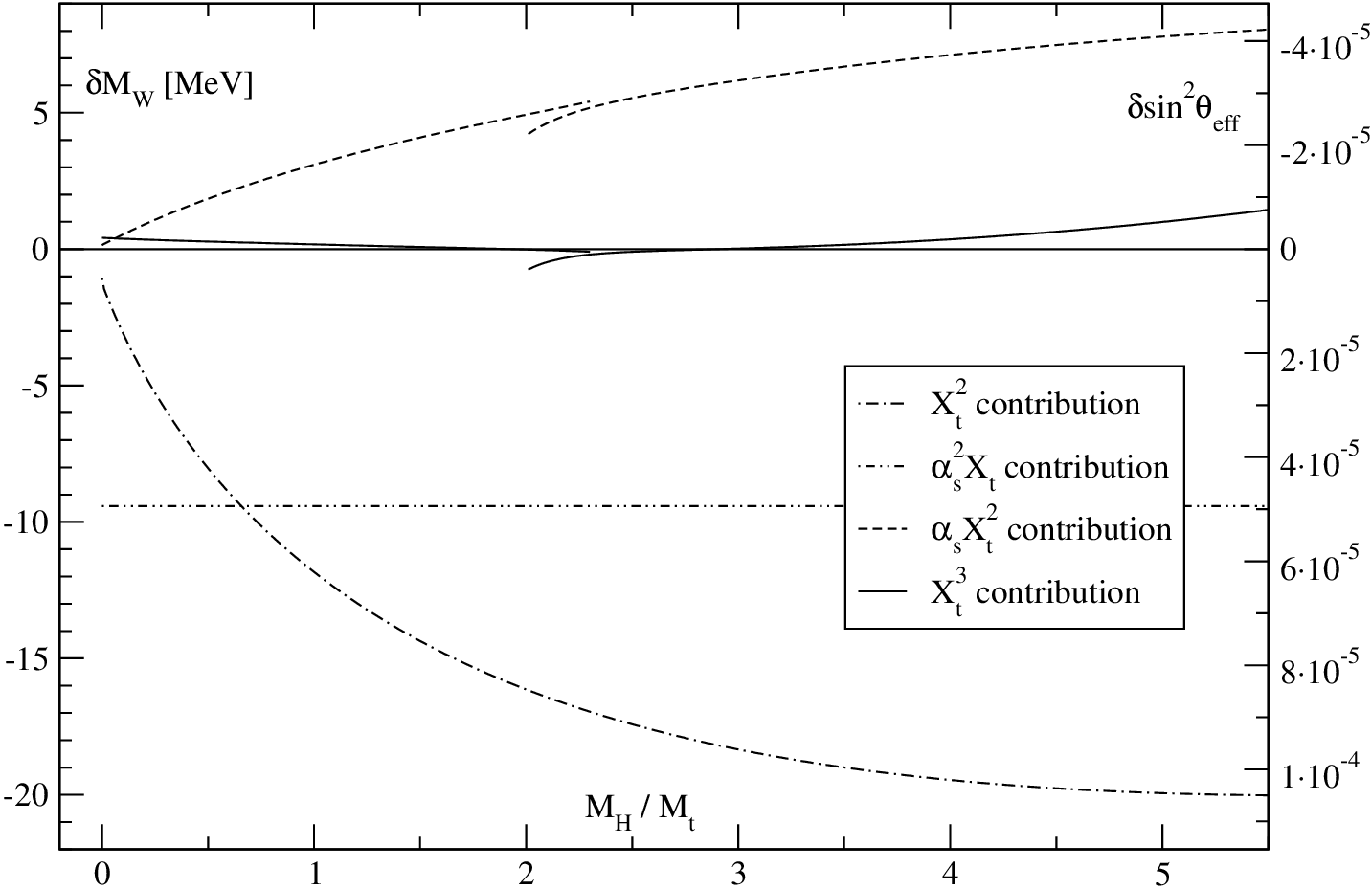}
}
\caption{Shifts in $\MW$, $\sweff$ from the ${\cal O}(\GF^3\mt^6)$
(labelled ``$X_{\rm t}^3$ contribution'') and
${\cal O}(\GF^2\alps\mt^4)$ 
(labelled ``$\alps X_{\rm t}^2$ contribution'') contributions to $\De\rho$ (from
\citere{faisst}).
}
\label{fig:rho3l_kuehn_shifts}
\end{figure}

Beyond two-loop order the results for the pure fermion-loop
corrections (i.e.\ contributions containing $n$ fermion loops at
$n$-loop order) are known up to four-loop order~\cite{floops}. They
contain in particular the leading contributions in $\De\al$ and
$\De\rho$. Most recently results for the leading three-loop contributions of 
${\cal O}(\GF^3 \Mt^6)$ and ${\cal O}(\GF^2 \alps \Mt^4)$ to the $\rho$
parameter,
\beq
\De\rho^{(3)} = \frac{\Si_{\rm Z}^{(3)}(0)}{\MZ^2}
              - \frac{\Si_{\rm W}^{(3)}(0)}{\MW^2}
\label{eq:delrho}
\eeq
have been
obtained for arbitrary values of $\MH$ (by means of expansions around 
$\MH = \mt$ and for $\MH \gg \mt$)~\cite{faisst}, generalising a
previous result which was obtained in the limit $\MH = 0$~\cite{faisstold}.
In \refeq{eq:delrho} $\Si_{\rm Z}^{(3)}(0)$ and $\Si_{\rm W}^{(3)}(0)$
denote the ${\cal O}(\GF^3 \Mt^6)$ and ${\cal O}(\GF^2 \alps \Mt^4)$
contributions to the transverse parts of the Z~and W~self-energies at
vanishing external momentum. The corresponding shifts in $\MW$ and
$\sweff$ are given by
\beqar
\De\MW^{(3)} &\approx& \frac{\MW}{2} \frac{\cw^2}{\cw^2 - \sw^2}
\De\rho^{(3)}, \non \\
\De\sweff^{(3)} &\approx& -\,\frac{\cw^2 \sw^2}{\cw^2 - \sw^2}
\De\rho^{(3)} .
\label{eq:delrhoshifts}
\eeqar
Their numerical effect is shown in \reffi{fig:rho3l_kuehn_shifts}. The
${\cal O}(\GF^2 \alps \Mt^4)$ contributions lead to a shift in $\MW$ of
up to 5~MeV and in $\sweff$ of up to $2.5 \times 10^{-5}$ for $\MH \lsim
350$~GeV. The effect of the ${\cal O}(\GF^3 \Mt^6)$ contributions, on
the other hand, is small. It does not exceed 1~MeV and $1 \times
10^{-5}$ for $\MH \lsim 1$~TeV.

While for $\MW$ the complete electroweak two-loop result is known, the
prediction for $\sweff$ is currently based at the two-loop level 
on an expansion for large $\mt$ up to the next-to-leading term of 
${\cal O}(\GF^2\mt^2\MZ^2)$~\cite{dgs}. An evaluation of the complete
two-loop contributions to $\sweff$ is in progress~\cite{swefffull}.


\section{Simple parametrisation of the full result for the W-boson mass}

The full result for $\MW$ containing all relevant corrections known so
far is obtained from $\De r$ given by
\beqar
\De r &=& \De r^{(\al)} + \De r^{(\al\alps)} + \De r^{(\al\alps^2)} +
\De r^{(\al^2)}_{\rm ferm} + \De r^{(\al^2)}_{\rm bos} \non \\
&& {} + \De r^{(\GF^2 \alps \Mt^4)} + \De r^{(\GF^3 \Mt^6)} ,
\label{eq:delrcontribs}
\eeqar
where $\De r^{(\al)}$ is the one-loop result, \refeq{eq:delrol}, $\De
r^{(\al\alps)}$ and $\De r^{(\al\alps^2)}$ are the two-loop~\cite{qcd2}
and three-loop~\cite{qcd3,qcd3light} QCD corrections, and 
$\De r^{(\al^2)}_{\rm ferm}$~\cite{2lferm,2lfermb,2lfermc} and 
$\De r^{(\al^2)}_{\rm bos}$~\cite{2lbos} are the fermionic and purely 
bosonic electroweak two-loop corrections, respectively. 
The contributions $\De r^{(\GF^2 \alps \Mt^4)}$ and 
$\De r^{(\GF^3 \Mt^6)}$ are obtained from the leading three-loop
corrections to $\De\rho$~\cite{faisst} specified in \refeq{eq:delrho}.

In \refeq{eq:delrcontribs}
the pure fermion-loop contributions at three-loop
and four-loop order obtained in \citere{floops} are not included 
because their
contribution turned out to be small as a consequence of accidental
numerical cancellations, with a net effect of only about 1~MeV in $\MW$
(using the real-pole definition of the gauge-boson masses). Since the
result given in \citere{floops} contains the leading contributions
involving powers of $\De\al$ and $\De\rho$ beyond two-loop order, it is
not necessary to
make use of resummations of $\De\al$ and $\De\rho$ as it was often
done in the literature in the past (see e.g.\ \citeres{resum}).
Accordingly, the quantity $\De r$ appears in \refeq{eq:delr} in fully
expanded form.

In \refta{tab:delrcontribs} the numerical values of the different
contributions to $\De r$ are given for $\MW =
80.426$~GeV~\cite{datasummer2003}. The other input parameters 
are~\cite{datasummer2003}
\beqar
&& \mt = 174.3 \gev, \; \mb = 4.7 \gev, \non \\
&& \MZ = 91.1875 \gev, \; \Ga_{\PZ} = 2.4952 \gev, \non \\
&& \al^{-1} = 137.03599976, \;
\De\al = 0.05907, \; \alps(\MZ) = 0.119, \non \\
&& \GF = 1.16637 \times 10^{-5} \gev^{-2} , 
\label{eq:inputs}
\eeqar
where $\De\al \equiv \De\al_{\rm lept} + \De\al^{(5)}_{\rm had}$.
The total width of the Z~boson, $\Ga_{\PZ}$, appears as an input parameter 
since the experimental value of $\MZ$ in \refeq{eq:inputs}, corresponding to a
Breit--Wigner parametrisation with running width, needs to be
transformed into the mass parameter defined according
to the real part of the complex pole, which corresponds to a
Breit--Wigner parametrisation with a constant decay width, see
\citere{2lfermb}. It is understood that $\MW$ in this paper always
refers to the conventional definition according to a Breit--Wigner
parametrisation with running width. The change of parametrisation is
achieved with the one loop QCD corrected value of the W-boson width as
described in \citere{2lfermb}.

\begin{table*}[tp]
$$
\begin{array}{|c||c|c|c|c|c|c|c|} \hline
\MH /\GeV &
\De r^{(\al)} & \De r^{(\al\alps)} & \De r^{(\al\alps^2)} &
\De r^{(\al^2)}_{\rm ferm} & \De r^{(\al^2)}_{\rm bos}  &
\De r^{(\GF^2 \alps \Mt^4)} & \De r^{(\GF^3 \Mt^6)}
\\ \hline
100  & 283.41 & 35.89 & 7.23 & 28.56 & 0.64  & -1.27 & -0.16 \\
200  & 307.35 & 35.89 & 7.23 & 30.02 & 0.35  & -2.11 & -0.09 \\
300  & 323.27 & 35.89 & 7.23 & 31.10 & 0.23  & -2.77 & -0.03 \\
600  & 353.01 & 35.89 & 7.23 & 32.68 & 0.05  & -4.10 & -0.09 \\
1000 & 376.27 & 35.89 & 7.23 & 32.36 & -0.41 & -5.04 & -1.04 \\ \hline
\end{array}
$$
\caption{The numerical values ($\times 10^4$) of the different 
contributions to $\De r$ specified in \refeq{tab:delrcontribs} are given 
for different values of $\MH$ and 
$\MW = 80.426$~GeV (the W and Z masses have been transformed so as to
correspond to the real part of the complex pole). The other input
parameters are listed in \refeq{eq:inputs} (from \citere{mw2loop}).
\label{tab:delrcontribs}}
\end{table*}

\refta{tab:delrcontribs} shows that the two-loop QCD correction,\\ 
$\De r^{(\al\alps)}$, and the fermionic electroweak two-loop correction,
$\De r^{(\al^2)}_{\rm ferm}$ are of similar size. They both amount to
about 10\% of the one-loop contribution, $\De r^{(\al)}$, 
entering with the same sign.
The most important correction beyond these contributions is the
three-loop QCD correction, $\De r^{(\al\alps^2)}$, which leads to a
shift in $\MW$ of about $-11$~MeV. For large values of $\MH$ also the 
contribution $\De r^{(\GF^2 \alps \Mt^4)}$ becomes sizable (see 
also the discussion of \reffi{fig:rho3l_kuehn_shifts}). The purely 
bosonic two-loop contribution, $\De r^{(\al^2)}_{\rm bos}$, and the leading
electroweak three-loop correction,\\
$\De r^{(\GF^3 \Mt^6)}$, give rise to
shifts in $\MW$ which are much smaller than even the experimental
error envisaged for a future Linear Collider, 
$\de\MW^{\rm exp, LC} = 7$~MeV~\cite{mtmwlc}. 

Since $\De r$ is evaluated in \refta{tab:delrcontribs} for a fixed value 
of $\MW$, the contributions $\De r^{(\al\alps)}$ and $\De r^{(\al\alps^2)}$
are $\MH$-in\-de\-pen\-dent. In the iterative procedure for evaluating $\MW$
from $\De r$, on the other hand, also these contributions
become $\MH$-dependent through the $\MH$-dependence of the inserted
$\MW$ value.

The electroweak two-loop result for $\MW$ is very lengthy and involves
numerical integrations of two-loop scalar integrals. It is therefore not
possible to present the result for $\MW$ in a compact analytic form. 
Instead, the full 
result for $\MW$, incorporating all corrections listed in 
\refeq{eq:delrcontribs}, can be approximated by the following simple
parametrisation~\cite{mw2loop},
\beqar
\MW &=& \MW^0 - c_1 \, \mathrm{dH} - c_2 \, \mathrm{dH}^2 
       + c_3 \, \mathrm{dH}^4 + c_4 (\mathrm{dh} - 1) \non \\
&& {} - c_5 \, \mathrm{d}\al + c_6 \, \mathrm{dt} 
       - c_7 \, \mathrm{dt}^2
       - c_8 \, \mathrm{dH} \, \mathrm{dt} 
       + c_9 \, \mathrm{dh} \, \mathrm{dt} \non \\
&& {} - c_{10} \, \mathrm{d}\alps
       + c_{11} \, \mathrm{dZ} ,
\label{eq:fitformula}
\eeqar
where
\beqar
&& \mathrm{dH} = \ln\left(\frac{\MH}{100 \gev}\right), \;
\mathrm{dh} = \left(\frac{\MH}{100 \gev}\right)^2, \non \\
&& \mathrm{dt} = \left(\frac{\mt}{174.3 \gev}\right)^2 - 1, \;
\mathrm{dZ} = \frac{\MZ}{91.1875 \gev} -1, \non \\
&& \mathrm{d}\al = \frac{\De\al}{0.05907} - 1, \;
\mathrm{d}\alps = \frac{\alps(\MZ)}{0.119} - 1 ,
\label{eq:pardef}
\eeqar
and the coefficients $\MW^0, c_1, \ldots, c_{11}$ take the following values
(in GeV)
\beqar
\MW^0 = 80.3799 , & c_1 = 0.05429 , & c_2 = 0.008939  , \non \\
c_3 = 0.0000890 , & c_4 = 0.000161 , & c_5 = 1.070  , \non \\
c_6 = 0.5256 , & c_7 = 0.0678 , & c_8 = 0.00179  , \non \\
c_9 = 0.0000659 , & c_{10} = 0.0737 , & c_{11} = 114.9  .
\label{eq:fitparams}
\eeqar
The parametrisation given in
\refeqs{eq:fitformula}--(\ref{eq:fitparams}) approximates the full
result for $\MW$ to better than 0.5~MeV over the whole range of 
$10 \gev \leq \MH \leq 1 \tev$ if all other experimental input values
vary within their combined $2 \si$ region around their central values 
given in \refeq{eq:pardef}. This should be sufficiently accurate for 
practical applications. 

In view of the experimental exclusion bound on
the Higgs-boson mass of $\MH > 114.4$~GeV~\cite{mhlimit} it seems
reasonable to restrict the Higgs-boson mass to the range $100 \gev \leq
\MH \leq 1 \tev$. In this case a slight readjustment of the coefficients
in \refeq{eq:fitparams} yields a parametrisation which approximates the
full result for $\MW$ even within 0.2~MeV, see \citere{mw2loop}.


\section{Remaining theoretical uncertainties}

The theoretical predictions for the electroweak precision observables
are affected by two kinds of uncertainties, namely the parametric 
uncertainty induced by the experimental errors of the input parameters,
e.g.\ $\mt$, and the uncertainty from unknown higher-order
corrections.  

\btab[h]
$$
\begin{array}{|c||c|c|} \hline
 & \de\MW/\mev & \de\sweff/10^{-5}
\\ \hline
\de\mt = 5.1 \gev & 31 & -16 \\
\de\MZ = 2.1 \mev & 2.6 & 1.4 \\
\de\left(\De\al^{(5)}_{\rm had}\right) = 0.00036 & -6.5 & 13 \\
\de\alps(\MZ) = 0.0027 & -1.7 & 1.0 \\
\hline
\end{array}
$$
\caption{Approximate shifts in $\MW$ and $\sweff$ caused by varying the
input parameters $\mt$, $\MZ$, $\De\al^{(5)}_{\rm had}$ and $\alps(\MZ)$
by $1 \si$ around their experimental central
values~\cite{datasummer2003}. 
\label{tab:paramunc}}
\etab

The parametric uncertainties induced by varying the input values of
$\mt$, $\MZ$, $\De\al^{(5)}_{\rm had}$ and $\alps(\MZ)$ by one standard
deviation are shown for $\MW$ and $\sweff$ in \refta{tab:paramunc}.
The dominant parametric 
uncertainty at present (besides the dependence on $\MH$) 
is induced by the experimental error of the
top-quark mass. It is about as large as the current experimental error
for both $\MW$ and $\sweff$.
The uncertainty caused by the experimental error of $\mt$ will remain
the dominant source of theoretical uncertainty in the prediction for
$\MW$ and $\sweff$ even at the LHC, where the error on $\mt$ will be 
reduced to $\de\mt = 1$--2~GeV~\cite{mtlhc}. A further improvement of the
parametric uncertainty of $\MW$ will require the precise measurement of 
$\mt$ at a future Linear Collider~\cite{delmt}, where an accuracy
of about $\de\mt = 0.1$~GeV will be achievable~\cite{mtmwlc}. 

The second source of theoretical uncertainties in the
prediction of the electroweak precision observables are 
the uncertainties from unknown higher-order 
corrections. Different approaches have been used in the literature for
estimating the possible size of uncalculated higher-order 
corrections, see e.g.~\citeres{mwest,2lfermb}. 
Since several of the corrections whose possible size had been estimated
in the past have meanwhile been calculated, there exists some guidance
concerning the reliability of the different methods. In \citere{mw2loop}
a careful analysis of the remaining uncertainties from unknown
higher-order corrections in the prediction for $\MW$ has been carried
out. The three main sources of uncertainties in the prediction of $\MW$
are from uncalculated corrections at ${\cal O}(\GF^2 \alps \Mt^2\MZ^2)$,
${\cal O}(\GF^3 \Mt^4 \MZ^2)$ and ${\cal O}(\al\alps^3)$. The resulting
theoretical uncertainty in the prediction for $\MW$ has been estimated
in \citere{mw2loop} to be
\beq
\de\MW^{\rm theo} \approx 4 \mev.
\label{eq:mwtheounc}
\eeq
This estimate holds for a relatively light Higgs boson, $\MH \lsim
300$~GeV. For a heavy Higgs boson, i.e.\ $\MH$ close to the TeV scale, the 
remaining theoretical uncertainty is significantly larger.

While for the case of $\MW$ unknown higher-order corrections are
encountered only beyond the two-loop level, the prediction for $\sweff$ is
affected by further uncertainties arising from the non-leading fermionic
two-loop contributions and the purely bosonic two-loop contributions, 
which have not yet been calculated. Using the same methods for
estimating the theoretical uncertainties as in \citere{mw2loop}, one finds
for the remaining theoretical uncertainty in the prediction for $\sweff$
from unknown higher-order corrections
\beq
\de\sweff^{\rm theo} \approx 6 \times 10^{-5} .
\label{eq:swefftheounc}
\eeq
The theoretical uncertainty of $\sweff$ is the dominant contribution to
the ``Blue Band'' indicating the effect of the theoretical uncertainties
from unknown higher-order corrections in the global SM fit to all
data~\cite{datasummer2003,quast03}.


\section{Comparison of the SM prediction for $\MW$ with the experimental
result}

\begin{figure}[ht]
\resizebox{0.5\textwidth}{!}{%
  \includegraphics{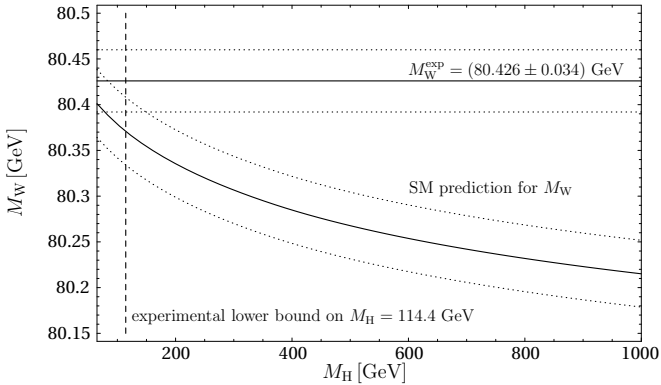}
}
\caption{Prediction for $\MW$ in the SM as a function of $\MH$ for $\mt
= 174.3 \pm 5.1$~GeV. The current experimental value,
$\MW^{\mathrm{exp}} = 80.426 \pm 0.034$~GeV~\cite{datasummer2003}, and
the experimental 95\% C.L.\ lower bound on the Higgs-boson mass,
$\MH = 114.4$~GeV~\cite{mhlimit}, are also indicated (from
\citere{mw2loop}).
}
\label{fig:MWpred}
\end{figure}

The theoretical prediction for $\MW$ within the SM is shown as a
function of the Higgs-boson mass in \reffi{fig:MWpred}. The width of the
band indicates the theoretical uncertainties, which contain the
parametric uncertainties from varying the input parameters within one
standard deviation (see \refta{tab:paramunc}) and the estimate of the 
uncertainties from unknown higher-order corrections given in
\refeq{eq:mwtheounc}. As discussed above, the theoretical uncertainty is
dominated by the effect of the experimental error of the top-quark mass.

The theoretical prediction is compared in
\reffi{fig:MWpred} with the current experimental
value~\cite{datasummer2003}, taking into account the 95\% exclusion
bound from the direct search for the SM Higgs, 
$\MH > 114.4$~GeV~\cite{mhlimit}.
The comparison clearly favours a light Higgs-boson mass within the SM.
Above the LEP exclusion bound on $\MH$ 
the $1 \si$ bands of the theory prediction and the
experimental result for $\MW$ overlap only in a small region,
corresponding to $\MH$ values significantly below 200~GeV.


\subsection*{Acknowledgements}

The author thanks M.~Awramik, M.~Czakon and A.~Frei\-tas for collaboration
on some of the results discussed in this paper.




\begin{thebibliography}{}

\bibitem{datasummer2003}
P.~Wells, talk presented at HEP2003 Europhysics Conference, Aachen, July
2003, to appear in the proceedings.

\bibitem{sirlin}
A.~Sirlin,
Phys.\ Rev.\ D {\bf 22} (1980) 971;\\
W.~J.~Marciano and A.~Sirlin,
Phys.\ Rev.\ D {\bf 22} (1980) 2695
[Erratum-ibid.\ D {\bf 31} (1985) 213].

\bibitem{velt}
M.~J.~Veltman,
Nucl.\ Phys.\ B {\bf 123} (1977) 89.

\bibitem{qcd2}
A.~Djouadi and C.~Verzegnassi,
Phys.\ Lett.\ B {\bf 195} (1987) 265;\\
A.~Djouadi,
Nuovo Cim.\ A {\bf 100} (1988) 357;\\
B.~A.~Kniehl,
Nucl.\ Phys.\ B {\bf 347} (1990) 86;\\
F.~Halzen and B.~A.~Kniehl,
Nucl.\ Phys.\ B {\bf 353} (1991) 567;\\
B.~A.~Kniehl and A.~Sirlin,
Nucl.\ Phys.\ B {\bf 371} (1992) 141;\\
B.~A.~Kniehl and A.~Sirlin,
Phys.\ Rev.\ D {\bf 47} (1993) 883;\\
A.~Djouadi and P.~Gambino,
Phys.\ Rev.\ D {\bf 49} (1994) 3499
[Erratum-ibid.\ D {\bf 53} (1994) 4111]
[arXiv:hep-ph/9309298].

\bibitem{qcd3}
L.~Avdeev, J.~Fleischer, S.~Mikhailov and O.~Tarasov,
Phys.\ Lett.\ B {\bf 336} (1994) 560
[Erratum-ibid.\ B {\bf 349} (1994) 597]
[arXiv:hep-ph/9406363];\\
K.~G.~Chetyrkin, J.~H.~K\"uhn and M.~Steinhauser,
Phys.\ Lett.\ B {\bf 351} (1995) 331
[arXiv:hep-ph/9502291];\\
K.~G.~Chetyrkin, J.~H.~K\"uhn and M.~Steinhauser,
Phys.\ Rev.\ Lett.\  {\bf 75} (1995) 3394
[arXiv:hep-ph/9504413].

\bibitem{qcd3light}
K.~G.~Chetyrkin, J.~H.~K\"uhn and M.~Steinhauser,
Nucl.\ Phys.\ B {\bf 482} (1996) 213
[arXiv:hep-ph/9606230].

\bibitem{2lferm} 
A.~Freitas, W.~Hollik, W.~Walter and G.~Weiglein,
Phys.\ Lett.\ B {\bf 495} (2000) 338
[Erratum-ibid.\ B {\bf 570} (2003) 260]
[arXiv:hep-ph/0007091].

\bibitem{2lfermb}
A.~Freitas, W.~Hollik, W.~Walter and G.~Weiglein,
Nucl.\ Phys.\ B {\bf 632} (2002) 189
[Erratum-ibid.\ B {\bf 666} (2003) 305]
[arXiv:hep-ph/0202131].

\bibitem{2lfermc}
M.~Awramik and M.~Czakon,
Phys.\ Lett.\ B {\bf 568} (2003) 48
[arXiv:hep-ph/0305248].

\bibitem{2lbos}
M.~Awramik and M.~Czakon,
Phys.\ Rev.\ Lett.\  {\bf 89} (2002) 241801
[arXiv:hep-ph/0208113]; 
{\it see also}
Nucl.\ Phys.\ Proc.\ Suppl.\  {\bf 116} (2003) 238
[arXiv:hep-ph/0211041].\\
A.~Onishchenko and O.~Veretin,
Phys.\ Lett.\ B {\bf 551} (2003) 111
[arXiv:hep-ph/0209010];\\
M.~Awramik, M.~Czakon, A.~Onishchenko and O.~Veretin,
Phys.\ Rev.\ D {\bf 68} (2003) 053004
[arXiv:hep-ph/0209084].

\bibitem{floops}
G.~Weiglein,
Acta Phys.\ Polon.\ B {\bf 29} (1998) 2735
[hep-ph/9807222];\\
A.~Stremplat, Diploma thesis (Univ.\ of Karlsruhe, 1998).

\bibitem{faisst}
M.~Faisst, J.~H.~K\"uhn, T.~Seidensticker and O.~Veretin,
Nucl.\ Phys.\ B {\bf 665} (2003) 649
[arXiv:hep-ph/0302275].

\bibitem{faisstold}
J.~J.~van der Bij, K.~G.~Chetyrkin, M.~Faisst, G.~Jikia and
T.~Seidensticker,
Phys.\ Lett.\ B {\bf 498} (2001) 156
[arXiv:hep-ph/0011373].

\bibitem{dgs}
G.~Degrassi, P.~Gambino and A.~Sirlin,
Phys.\ Lett.\ B {\bf 394} (1997) 188
[arXiv:hep-ph/9611363].

\bibitem{swefffull}
M.~Awramik, M.~Czakon, A.~Freitas and G.~Weiglein, work in progress;\\
M.~Czakon, talk given at the ECFA Linear Collider Workshop,
Montpellier, November 2003.


\bibitem{resum}
W.~J.~Marciano,
Phys.\ Rev.\ D {\bf 20} (1979) 274;\\
A.~Sirlin,
Phys.\ Rev.\ D {\bf 29} (1984) 89;\\
M.~Consoli, W.~Hollik and F.~Jegerlehner,
Phys.\ Lett.\ B {\bf 227} (1989) 167.

\bibitem{mw2loop}
M.~Awramik, M.~Czakon, A.~Freitas and G.~Weiglein,
arXiv:hep-ph/0311148.

\bibitem{mtmwlc}
J.~A.~Aguilar-Saavedra {\it et al.}  [ECFA/DESY LC Physics Working Group
                  Collaboration],
arXiv:hep-ph/0106315;\\
T.~Abe {\it et al.}  [American Linear Collider Working Group
Collaboration],
in {\it Proc. of the APS/DPF/DPB Summer Study on the Future of Particle
Physics (Snowmass 2001) } ed. N.~Graf,
arXiv:hep-ex/0106055;\\
K.~Abe {\it et al.}  [ACFA Linear Collider Working Group Collaboration],
arXiv:hep-ph/0109166;
see: {\tt lcdev.kek.jp/RMdraft/}~.

\bibitem{mhlimit}
[The LEP working group for Higgs boson searches],
Phys.\ Lett.\ B {\bf 565} (2003) 61
[arXiv:hep-ex/0306033].

\bibitem{mtlhc}
M.~Beneke {\it et al.},
arXiv:hep-ph/0003033,
in: Standard Model Physics (and more) at the LHC,
eds.\ G.~Altarelli and M.~Mangano,
CERN, Geneva, 1999 [CERN-2000-004].

\bibitem{delmt}
S.~Heinemeyer, S.~Kraml, W.~Porod and G.~Weiglein,
JHEP {\bf 0309} (2003) 075
[arXiv:hep-ph/0306181].

\bibitem{mwest}
D.~Y.~Bardin {\it et al.},
arXiv:hep-ph/9709229;\\
D.~Y.~Bardin, M.~Grunewald and G.~Passarino,
arXiv:hep-ph/9902452;\\
P.~Gambino,
arXiv:hep-ph/9812332;\\
A.~Freitas, S.~Heinemeyer, W.~Hollik, W.~Walter and G.~Weiglein,
Nucl.\ Phys.\ Proc.\ Suppl.\  {\bf 89} (2000) 82
[arXiv:hep-ph/0007129];\\
A.~Ferroglia, G.~Ossola and A.~Sirlin,
Phys.\ Lett.\ B {\bf 507} (2001) 147
[arXiv:hep-ph/0103001];\\
U.~Baur {\it et al.},
hep-ph/0202001,
in {\it Proc. of the APS/DPF/DPB Summer Study on the Future of Particle
Physics (Snowmass 2001) } eds. R.~Davidson and C.~Quigg;\\
A.~Freitas, S.~Heinemeyer and G.~Weiglein,
Nucl.\ Phys.\ Proc.\ Suppl.\  {\bf 116} (2003) 331
[arXiv:hep-ph/0212068].

\bibitem{quast03}
G.~Quast, talk presented at HEP2003 Europhysics Conference, Aachen, July
2003, to appear in the proceedings.

\end{thebibliography}
\end{document}